\documentclass[fleqn,usenatbib]{rasti}

\usepackage{newtxtext,newtxmath}
\usepackage[T1]{fontenc}
\usepackage{ae,aecompl}
\usepackage{bm}
\usepackage{orcidlink}

\usepackage{graphicx}	
\usepackage{amsmath}	
\title[RAMSES-GPU: Adaptive Mesh Refinement on the GPU]{RAMSES-GPU: Cell-by-Cell Adaptive Mesh Refinement with Magneto-Hydrodynamics and Self-Gravity on Graphics Processing Units}

\author[R. Teyssier et al.]{%
Romain Teyssier\,\orcidlink{0000-0001-7689-0933}$^{1}$,\thanks{E-mail: teyssier@princeton.edu}
Eric Moseley\,\orcidlink{0000-0001-8558-5009}$^{2}$,
Robert V. Caddy\,\orcidlink{0000-0002-4475-3181}$^{5}$,
Troels Haugb\o{}lle\,\orcidlink{0000-0002-9422-8684}$^{4}$,
James Sunseri\,\orcidlink{0000-0003-4274-2662}$^{1}$,
\newauthor
Jun-Young Lee\,\orcidlink{0009-0006-4981-0604}$^{1}$,
Jonathan Palafoutas\,\orcidlink{0000-0002-3411-8897}$^{1}$,
Yue Pan\,\orcidlink{0000-0002-7922-9726}$^{1}$,
William Groger\,\orcidlink{0009-0009-0500-4514}$^{1}$,
Tom Abel\,\orcidlink{0000-0002-5969-1251}$^{2}$,
\newauthor
Burlen Loring\,\orcidlink{0000-0002-4678-8142}$^{3}$,
Guillaume Barnier\,$^{3}$
\\
$^{1}$Department of Astrophysical Sciences, Princeton University, Princeton, NJ 08544, USA\\
$^{2}$Kavli Institute for Particle Astrophysics and Cosmology, Stanford University, Stanford, CA 94305, USA\\
$^{3}$NVIDIA Corporation, Santa Clara, CA 95051, USA\\
$^{4}$Niels Bohr Institute, University of Copenhagen, Blegdamsvej 17, 2100 Copenhagen, Denmark\\
$^{5}$Princeton Research Computing, Princeton University, Princeton, NJ 08544, USA
}

\date{Accepted XXX. Received YYY; in original form ZZZ}

\pubyear{2026}

\begin{document}
\label{firstpage}
\pagerange{\pageref{firstpage}--\pageref{lastpage}}
\maketitle

\begin{abstract}
We present the implementation and optimization of the cosmological simulation code \textsc{ramses} on Graphics Processing Units (GPUs) using CUDA Fortran. This accelerated version ports the main computational routines—including hydrodynamics, particle dynamics, and self-gravity—to multi-GPU architectures. We detail our strategy for managing cell-by-cell Adaptive Mesh Refinement (AMR) on the GPU, utilizing bucket sort with prefix sums for AMR level sorting, radix sort via the CUB library for Hilbert key ordering, and an \texttt{fnv64} hash table with linear probing for fast spatial indexing. Portability across diverse hardware architectures is achieved via a dispatcher and C-Fortran wrappers, calling CUDA, HIP, and Metal kernels directly translated from the CUDA Fortran framework. Hydrodynamics updates are executed via a Godunov MUSCL-Hancock HLLC Riemann solver managed through a three-tier shared-memory kernel architecture (named \texttt{rock}, \texttt{paper}, and \texttt{scissor}). Particle mass deposition uses Cloud-in-Cell (CIC) interpolation optimized with atomic additions or prefix sums, combined with a kick-drift-kick time integration pusher. Self-gravity is handled via a Multigrid (MG) Poisson solver performing hierarchical V-cycles on individual levels. Performance benchmarks conducted on NVIDIA A100 and H200 GPUs demonstrate substantial accelerations compared to multi-core CPUs, yielding $10\times$ up to a $100\times$ speedup for standard test problems such as the Sedov blast wave, molecular core collapse, and cosmological simulations. Finally, we briefly discuss additional accelerated physics modules, including equilibrium cooling, polytropic equations of state, ideal and non-ideal magneto-hydrodynamics (MHD), and stellar feedback.
\end{abstract}


\section{Introduction}

Numerical simulations are crucial tools in modern astrophysics and cosmology, enabling the study of non-linear structure formation from primordial fluctuations down to galactic and stellar scales. The code \textsc{ramses} \citep{Teyssier2002} has been widely adopted by the community due to its robust cell-by-cell Adaptive Mesh Refinement (AMR) capability, which allows for high spatial and mass resolution when physically warranted, such as in high-density regions of dark matter halos, interstellar media, and star-forming regions.

However, the rapid architectural evolution of modern High-Performance Computing (HPC) clusters toward heterogeneous systems heavily accelerated by Graphics Processing Units (GPUs) requires a fundamental redesign of traditional CPU-centric algorithms. Porting cell-by-cell AMR codes to GPUs presents significant challenges due to irregular tree data structures, sparse memory access patterns, and complex pointer-chasing operations inherent to adaptive grids. 

To address these hardware trends, several high-performance, mesh-based astrophysics codes have successfully shifted to or been rewritten for GPU architectures. These efforts can be broadly categorized on the basis of their underlying mesh paradigms:
\begin{itemize}
    \item \textbf{Block-Structured AMR Codes:} These use a patch- or block-based hierarchical mesh layout to dynamically vary resolution. Examples include \textsc{gamer} and \textsc{gamer-2} \citep{Schive2010,Schive2018}, which offload hydrodynamic, magnetohydrodynamic (MHD), and self-gravity solvers to the GPU; \textsc{quokka} \citep{Wibking2021}, a radiation-hydrodynamics code built on top of the \textsc{amrex} framework using a method-of-lines PPM fluid solver; \textsc{h-amr} \citep{Liska2022}, designed for general relativistic MHD (GRMHD) around black hole environments; and codes like \textsc{castro} \citep{Almgren2010} and \textsc{enzo-e} \citep{Bordner2018}, which have ported significant portions of their solvers using the \textsc{amrex} ecosystem \citep{Zhang2019}.
   
    \item \textbf{Cell-by-Cell AMR Codes:} While cell-by-cell adaptive tree structures are notoriously non-trivial to optimize for the wide-vector nature of GPUs, some targeted efforts have successfully ported fundamental hydrodynamics and MHD solvers of \textsc{ramses} to GPUs using dedicated OpenACC directives to offload heavy flux-evaluation routines. For example, \citet{Gheller2014} presented a comparative analysis of the GPUs porting strategies for \textsc{enzo} (using CUDA) and \textsc{ramses} (using OpenACC), highlighting the challenges of pointer-based tree traversals on GPUs and describing how compiler directives can offload flux evaluations.
     \item \textbf{Other AMR Codes:} These other codes  emphasize high-order accuracy schemes on structured or uniform grids. Some of them also leverage performance-portability layers like Kokkos \citep{Edwards2014}. Examples include \textsc{athenak} \citep{Stone2026}, a block-based AMR code natively rewritten using Kokkos to deploy Godunov methods, MHD, and GRMHD across different GPU vendors; \textsc{cholla}, a native uniform-grid code for GPU utilizing CUDA for large scale simulations that require high resolution throughout the simulation domain \citep{Schneider2015, Schneider2023, Caddy2024}; \textsc{gpluto}, a GPU-ready version of the widely used \textsc{pluto} \citep{Mignone2007} code utilizing OpenACC; and \textsc{idefix} \citep{Lesur2023}, which uses Kokkos for planetary disk and shearing-box simulations.
\end{itemize}

\begin{figure*}
    \centering
    \includegraphics[width=0.9\textwidth]{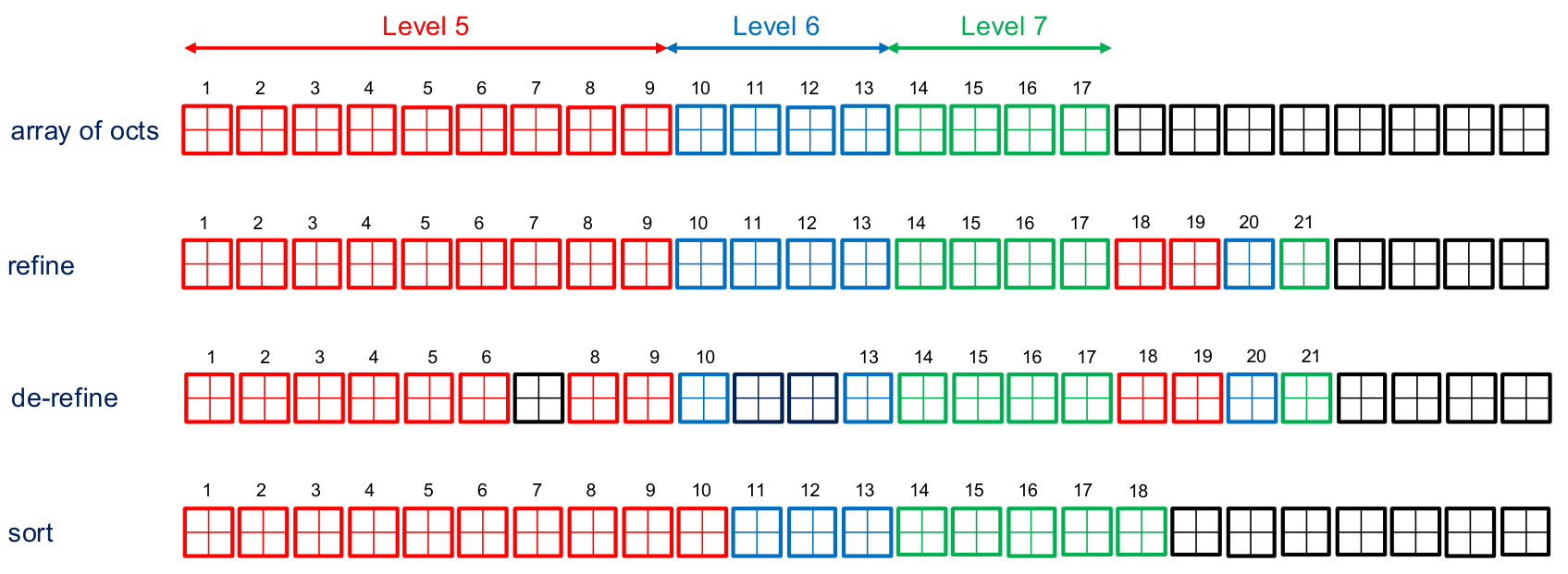}
    \caption{Schematic representation of the cell-by-cell Adaptive Mesh Refinement (AMR) structure on the GPU. Active oct blocks are level-sorted and ordered along a space-filling curve (Hilbert key) during refinement, de-refinement, and neighbor indexing to preserve memory locality and coalescing.}
    \label{fig:amr_schematic}
\end{figure*}

In addition to mesh-based solvers, cosmological $N$-body and particle-mesh/tree codes have also undergone extensive acceleration to exploit modern multi-GPU systems. Key implementations include \textsc{pkdgrav3} \citep{Potter2017}, a highly optimized tree code that enabled trillion-particle simulations for galaxy surveys; \textsc{abacus} \citep{Garrison2021}, which splits near-field and far-field gravitational forces to sustain massive performance for large-scale suites such as \textit{AbacusSummit} \citep{Maksimova2021}; \textsc{genga} and \textsc{genga ii} \citep{Grimm2014,Grimm2022}, GPU-native planetesimal codes employing hybrid symplectic integrators for planetary encounters; and \textsc{hacc} (Hardware/Hybrid Accelerated Cosmology Code) \citep{Habib2016}, which addresses heterogeneous architecture diversity at extreme scale, recently extended to exascale hydrodynamics via \textsc{crk-hacc} \citep{Frontiere2025}.

In this paper, we present the architecture and performance of \textsc{ramses-gpu}, an accelerated version of the code developed through a multi-institutional collaboration involving Princeton University (in particular support from the Princeton Research Software Engineer Program), Stanford University, the Niels Bohr Institute, and NVIDIA engineers under the auspices of several GPU Hackathons. Using CUDA Fortran alongside optimized GPU template libraries, we successfully ported the core solver routines (hydrodynamics, particles, and gravity) to modern GPU architectures, preserving the precise cell-by-cell resolution properties of the original framework. In contrast to the other efforts described previously for porting Ramses to GPUs, our implementation is based on a cell-by-cell AMR structure with level-by-level adaptive time stepping; and with data fully resident in device memory and modified in place using dedicated CUDA kernels.

\section{Adaptive Mesh Refinement on the GPU}

The cornerstone of the \textsc{ramses} framework is its graded-octree spatial discretization with cell-by-cell refinements and adaptive, level-by-level time steps. We use an array of \textit{octs} contiguously stored in GPU memory. The maximum array size is set at run time via the parameter \texttt{ngridmax}. In \textsc{ramses}, an oct is a small $2 \times 2 \times 2$ cell block and corresponds to the smallest block size if one aims at refining each cell individually. For performance optimization, in the following sections, we also explore grouping cells into larger blocks, typically $4 \times 4 \times 4$ or $8 \times 8 \times 8$ cell blocks. In the latter cases, the refinement strategy has to be adapted to enforce the desired block structure. 

\subsection{Hash Table Lookup}
In order to access neighboring cells in the AMR structure, we no longer use the Fully Threaded Tree structure \citep[][]{1998JCoPh.143..519K, Teyssier2002} and the associated pointer structure. We instead use a GPU-tailored hash table based on the \texttt{fnv64} hashing function with linear probing. Each oct is defined by a coordinate tuple $(i, j, k, l)$ that serves as the unique key, mapping directly to its position as a linear index in device memory. The key-value pairs are inserted into the hash table during refinement, and the value is set to 0 during the de-refinement. As a result, the hash table accumulates tombstones. These are removed by resetting the hash table and reinserting the entire AMR grid periodically (in practice, every coarse time step). For example, if we need access to the neighbor on the left, we just retrieve its index in memory from the hash table as \texttt{nbr\_idx = hash\_get(i-1,j,k,l)}. If we need access to its father index in memory, we use \texttt{father\_idx = hash\_get(i/2,j/2,k/2,l-1)}.

\subsection{Refinement and De-Refinement}

Our array of octs is sorted in an increasing level of refinements, starting from {\ttfamily ilevel=levelmin} up to {\ttfamily ilevel=nlevelmax}. Inside each level, octs are sorted along a 3D Hilbert index, statistically improving the memory locality of neighboring elements. At each time step, we modify the grid following 4 distinct stages (see Figure~\ref{fig:amr_schematic}):
\begin{itemize}
    \item \textbf{Refinement:} We create new oct blocks if a coarse cell is flagged for refinement. These new octs are inserted at the end of the oct array, and we set their level of refinement to their target level.
    \item \textbf{De-refinement:} If a split coarse cell is not flagged for refinement, we remove the child oct from memory. The oct array now has a hole. We set the corresponding level of refinement to zero.
    \item \textbf{Level sorting:} A bucket sort algorithm reorders the oct array level-by-level, putting the destroyed (level 0) octs at the end of the array.
    \item \textbf{Spatial ordering:} Within each level, octs are ordered along the Hilbert space-filling curve. We use a radix sort based on the NVIDIA CUB library \citep{Merrill2015, nvidias_cub}.
\end{itemize}

\subsection{Ghost Zones}

Although the algorithms described here work only with the hash table information, we found it more efficient to use father- and neighbor index arrays defined for every active oct. For stencil operations (hydro and gravity solvers), the neighbor index can point to a non-existent oct because it has not been refined. In this case, we create ghost zones dynamically using an array cache region. The size of this cache memory is set at run time via the parameter \texttt{ncachemax}. This cache bypasses completely hash table lookups and consolidates memory reads into continuous, high-bandwidth transfers at the expense of slightly more memory usage. This cache memory is built every time step right after the refinement/de-refinement step. Note that this cache memory will also be used to implement CUDA-aware MPI communications between multiple GPUs. This is left for future work.

\section{Hydrodynamics Solver}

\begin{figure*}
    \centering
    \includegraphics[width=\textwidth]{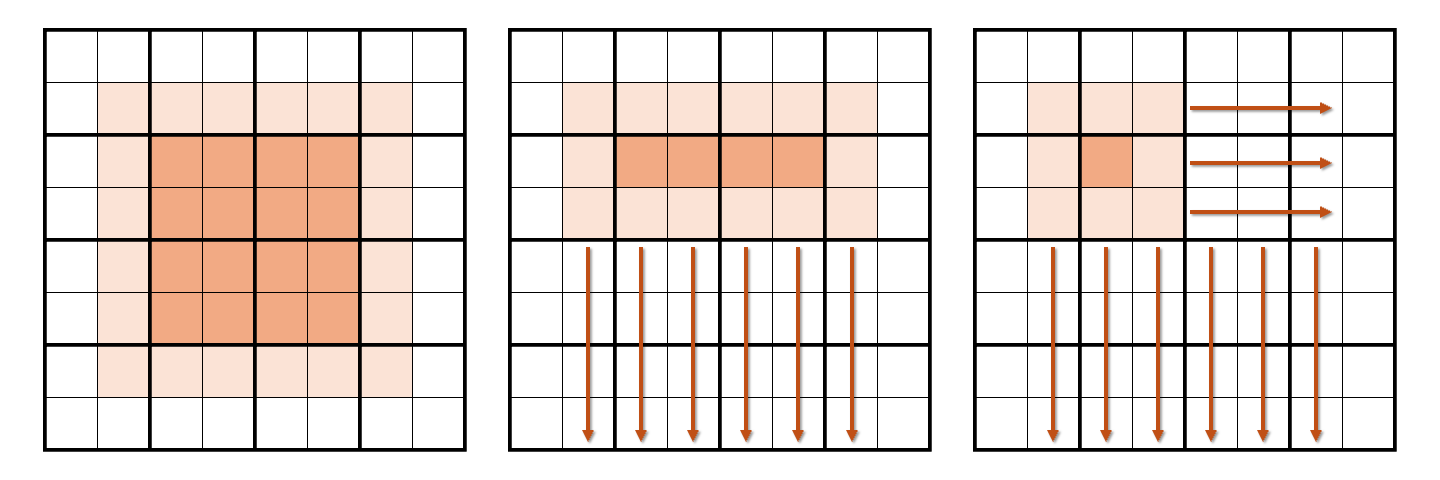}
    \caption{Shared memory footprint and data access layout for the three core hydrodynamics kernels ({\it rock}, {\it paper}, {\it scissor}) at {\ttfamily nsubgrid=2}. \textbf{Left}: The {\it rock} kernel caches 512 cells, including primitive variables and ghost zone halos. \textbf{Center}: The {\it paper} kernel operates on the reconstruction stencil of 192 cells with only 3 cells in the vertical direction. \textbf{Right}: The {\it scissor} kernel needs only 72 active cells but requires to sweep the cube both horizontally and vertically.}
    \label{fig:rps_kernels}
\end{figure*}

The hydrodynamics module solves the Euler or ideal MHD equations using a second-order Godunov MUSCL-Hancock scheme paired with an HLLC Riemann solver. This hydrodynamics solver \citep{Vanleer1979, Teyssier2002, Fromang2006} is based on a predictor-corrector, second-order time integration scheme and requires two ghost cells on each side of a cell to fully update the conservative variables. 

The general philosophy of our GPU implementation is to copy a block of $2 \times 2 \times 2$ cells (aka an oct) surrounded by two ghost cells in each direction into a local stencil stored in CUDA shared memory. This gives a local array of $6 \times 6 \times 6$ cells arranged in a cube that contains the old state of conservative variables, \texttt{uold}. The GPU kernel then works fully in shared memory to convert the conservative variables into primitive variables, reconstruct the second-order, face-centered left and right states, call the Riemann solver, and finally use the resulting fluxes to update the conservative variables, storing the result in the new state of conservative variables, \texttt{unew}.

The problem with this simple approach is the excessive redundant operations associated with the ghost cells. A more efficient approach consists of using a larger inner block of $2 \times 2 \times 2$ octs, with one ghost oct in each direction, so that the total shared
memory layout is based on $8 \times 8 \times 8$ cells, reducing the number of redundant operations. We define the number of octs in the inner block by the parameter \texttt{nsubgrid}, typically using values of 1, 2, or 4, resulting in a shared-memory block of $6 \times 6 \times 6$, $8 \times 8 \times 8$, or $12 \times 12 \times 12$. Figure~\ref{fig:rps_kernels} shows the example of \texttt{nsubgrid=2}. Increasing \texttt{nsubgrid} aggregates multiple oct blocks into larger spatial stencils, directly expanding the shared memory footprint per thread block while significantly increasing performance due to improved reuse of stencil data. 

However, the price to pay is a larger shared-memory footprint. For five conservative variables \texttt{nvar=5}, we need 12~kB of shared memory for \texttt{nsubgrid=1}, 41~kB for \texttt{nsubgrid=2}, and 210~kB for \texttt{nsubgrid=4}. However, modern GPUs default to enforcing a software limit of 48~kB of shared memory, which requires that at runtime we dynamically raise the shared memory limit before running our \texttt{nsubgrid = 4} kernel. Our \texttt{nsubgrid=4} kernel barely fits in the shared memory of the latest H200 NVIDIA GPUs, but not on older models like the A100. It is worth noting that while kernels with larger \texttt{nsubgrid} values do consume more shared memory, they use it more efficiently. The kernel \texttt{nsubgrid = 1} requires 1.2~kB of memory per real cell, while \texttt{nsubgrid = 2} and \texttt{nsubgrid = 4} require only 0.64~kB and 0.41~kB of shared memory per real cell, respectively.

We have also explored another option based on a planar stencil that sweeps the cube vertically from top to bottom. Because of the planar geometry of this kernel, we call it the \textit{paper kernel}, naturally leading to the name \textit{rock kernel} for the cubical kernel. This paper kernel is more complex, as it requires copying cells from the main GPU memory in a cyclic manner into a three-cell thick vertical stencil (see Figure~\ref{fig:rps_kernels}). However, it uses much less shared memory. 

The final option is the \textit{scissor kernel}, which relies on a pencil beam with a cross-section of $3 \times 3$ cells. It is the most efficient option for shared-memory, but the algorithm becomes significantly more complex as the kernel sweeps the cube both horizontally and vertically. Because of this extra complexity and the satisfactory performance of both the rock and paper kernels, we defer the implementation of the scissor kernel to future work.

In summary, we have three possible hydro kernels that operate as follows:
\begin{enumerate}
    \item \textbf{Rock Kernel:} It has a cubic shared-memory data layout with $6 \times 6 \times 6$ cells for \texttt{nsubgrid=1}, $8 \times 8 \times 8$ cells for \texttt{nsubgrid=2}, or $12 \times 12 \times 12$ cells for \texttt{nsubgrid=4}. For \texttt{nsubgrid=1}, this involves caching 216 cells ($12$~kB); for \texttt{nsubgrid=2}, this involves caching 512 cells ($40$~kB); and for \texttt{nsubgrid=4}, this involves caching 1728 cells ($204$~kB). In single precision, these sizes are roughly halved.
    \item \textbf{Paper Kernel:} It features a planar shared-memory data layout with $6 \times 6 \times 3$ cells for \texttt{nsubgrid=1}, $8 \times 8 \times 3$ cells for \texttt{nsubgrid=2}, or $12 \times 12 \times 3$ cells for \texttt{nsubgrid=4}. For \texttt{nsubgrid=1}, this involves caching 108 cells ($9$~kB); for \texttt{nsubgrid=2}, this involves caching 192 cells ($18$~kB); and for \texttt{nsubgrid=4}, this involves caching 432 cells ($45$~kB).
    \item \textbf{Scissor Kernel:} It features a shared-memory data layout with pencil-beams with $6 \times 3 \times 3$ cells for \texttt{nsubgrid = 1}, $8 \times 3 \times 3$ cells for \texttt{nsubgrid = 2}, or $12 \times 3 \times 3$ cells for \texttt{nsubgrid = 4}. We do not have estimates of its shared-memory footprint, as the kernel is yet to be developed.
\end{enumerate}

\section{Particle Solver}
 
 In this section, we describe the GPU port of the N-body solver in \textsc{ramses}. It is based on the Particle-Mesh (PM) method \citep{Hockney1981}, for which the particle masses are first deposited on the grid using the Cloud-in-Cell (CIC) deposition scheme \citep{BirdsallFuss1969}. The density is then used as the right-hand side of the Poisson equation, solved using our multigrid field solver (see Section~5). The potential is then converted into acceleration using a fourth-order finite difference approximation. The gravitational acceleration is then interpolated back to the particle positions using the inverse CIC scheme (i.e. tri-linear interpolation). Particle trajectories are finally integrated using the Verlet scheme, a variable time step variant of the leapfrog scheme (also known as a kick-drift-kick pusher). 

The particle positions, velocities, masses, refinement levels, and sorting arrays remain in the device memory throughout this sequence. If a particle's position falls within a given cell, that cell is called its \textit{source cell}. At each AMR level, we compute the Hilbert key of the particle source cells and radix-sort an index array. We use the CUB radix sort for the CUDA port and our in-house prefix-sum radix sort for our Metal port (more later). After depositing the particle masses, we use a second prefix sum to separate particles whose host cells are refined from those that remain at the current level. We then reorder the particle arrays and update the head and tail indices of each level. In this way, particles sharing a source cell remain contiguous during mass deposition while the particle list follows the AMR hierarchy from coarse to fine levels.

\subsection{Mass Deposition}
The most computationally intense segment involves mass deposition onto the underlying AMR grid. We employ a Cloud-in-Cell (CIC) interpolation scheme. To avoid race conditions when multiple particles concurrently deposit mass to the same grid cell, we utilize highly optimized atomic addition operations (\texttt{atomicAdd}) or parallel prefix sum kernels, depending on the deposition kernel selected at run time.

We have implemented three CIC deposition kernels, selected at run time with parameter \texttt{part\_dep\_algo=1, 2 or 3} corresponding to the {\it large}, {\it medium} or {\it small} kernel. These kernels show different performance for different particle distributions. The {\it large-stencil} kernel considers $3^3$ candidate cells around each particle's source cell, of which $2^3$ receive non-zero CIC weights. It combines particles belonging to the same source cell within a warp before performing the atomic additions. The main advantage of the {\it large} kernel is its simplicity, but we recommend using the two other kernels for higher performance. 

The default kernel (termed {\it medium} kernel) shifts the source-cell convention by half a cell, allowing us to consider only the $2^3$ cells with non-zero weights, and uses the same warp-level reduction. This kernel is based on extensive use of the AtomicAdd function. For that reason, it is well suited for particle distribution with few (up to 10-20) particles per cell. 

Finally, the {\it small} kernel uses the same shifted stencil, but replaces the atomic additions with prefix sums and one mesh update for each source-cell segment and stencil offset. It is ideally suited for particle distribution with more particles (say 1000) per cell.

At level $\ell$, each kernel deposits particles assigned to level $\ell$ and to all finer levels before the particle list is split for level $\ell+1$. All GPU kernels currently assume periodic boundaries.

\subsection{Force Interpolation}
The gravitational forces are interpolated from the grid back to the particle positions using the same Cloud-in-Cell (CIC) interpolation kernel executed in reverse.

For each particle, the force kernel first gathers the complete $2^3$ CIC stencil at the particle refinement level. If any cell in this stencil is unavailable, or belongs to a cache oct for which the force has not been updated, we discard the partial interpolation and repeat the complete CIC gather one level coarser. This treatment avoids mixing incomplete force stencils across a coarse--fine boundary.

\subsection{Particle Time Integration}

The particle pusher follows the recursive AMR time-step hierarchy. A kick-only operation applies one half-kick after the new gravitational force has been computed, while the complementary operation applies the other half-kick followed by a full drift. When a particle changes level, its stored level determines whether the old or new level time step is used for the remaining kick. For smoothly varying time steps this midpoint scheme is second-order accurate. A particle crossing a refinement boundary, however, encounters an abrupt factor-of-two change in time step, reducing the corresponding part of its trajectory to first-order accuracy \citep{Teyssier2002}.

\subsection{Current Scope}

The benchmarks presented here use dark matter on a single CUDA device. The CUDA implementation also contains the corresponding sorting, level-splitting, CIC, force-interpolation, and pusher kernels for star particles. Other particle types beyond dark matter and stars are not yet included in the CUDA implementation, although prototypes for sink particles, dust grains and cosmic rays are currently in development. The Metal implementation (more later) is presently restricted to dark matter, the medium CIC kernel, and single precision.

\section{Self-Gravity Solver}

In \textsc{ramses}, self-gravity is implemented via a Multigrid (MG) Poisson solver. The three main components of the algorithm have all been ported to their corresponding CUDA kernel: building the MG multi-resolution hierarchy, implementing the prolongation and restriction operators, and finally performing the Gauss-Seidel iterations. 

The Multigrid (MG) Poisson solver has been ported to
CUDA, replacing red-black Gauss--Seidel sweeps, restrictions,
prolongations, and residual computations on the CPU with a set of CUDA Fortran kernels in
the module \texttt{mg\_device}. All kernels use a two-dimensional thread block
of shape $(8, N_\text{threads})$ mapping \texttt{threadIdx\%x} to the eight
cells of an oct and \texttt{threadIdx\%y} to concurrent octs within the block,
so that one thread block processes $N_\text{threads}$ octs simultaneously.

The AMR neighbour topology is precomputed once per level into a flat device
array \texttt{nbor(1:27, *)} via hash-table lookups (\texttt{update\_nbor\_array\_mg}),
after which all per-kernel neighbor accesses are O(1) array dereferences,
replacing the MPI-aware hash-cache traversals of the CPU path. The
gravitational potential is built level by level using standard V-cycle
operations: the \texttt{restrict\_residual} kernel averages fine-level residuals
into the coarse RHS by full-weighting injection (factor $1/8$), and the
\texttt{interpolate\_correct} kernel prolongates corrections by trilinear CIC
interpolation with weights $(1, 3, 3, 9, 3, 9, 9, 27)/64$. The prolongation
kernel is the only MG kernel to use explicit shared memory, broadcasting the
27-parent-cell neighborhood to all eight cell-threads within a block to reduce
the number of hash lookups by a factor of eight.

The core smoother (\texttt{gauss\_seidel}) implements red-black
Gauss--Seidel, partitioning the eight cells of each oct into two sets of four
by the parity of their intra-oct coordinate index. The update formula
adjusts the effective diagonal of the discrete Laplacian to account for
partially cut boundary cells, using the signed distance function stored in
\texttt{f(:,3,:)} to enforce homogeneous Dirichlet conditions at the domain
boundary to second-order accuracy \citep{Guillet2011}. Global reductions (L2 residual norm,
maximum density) use a two-level warp-shuffle plus shared-memory block
reduction followed by device-level \texttt{atomicadd} and \texttt{atomicmax},
with the inter-node MPI reduction performed by the GPU runner layer.

\begin{figure}
    \centering
    \includegraphics[width=\linewidth]{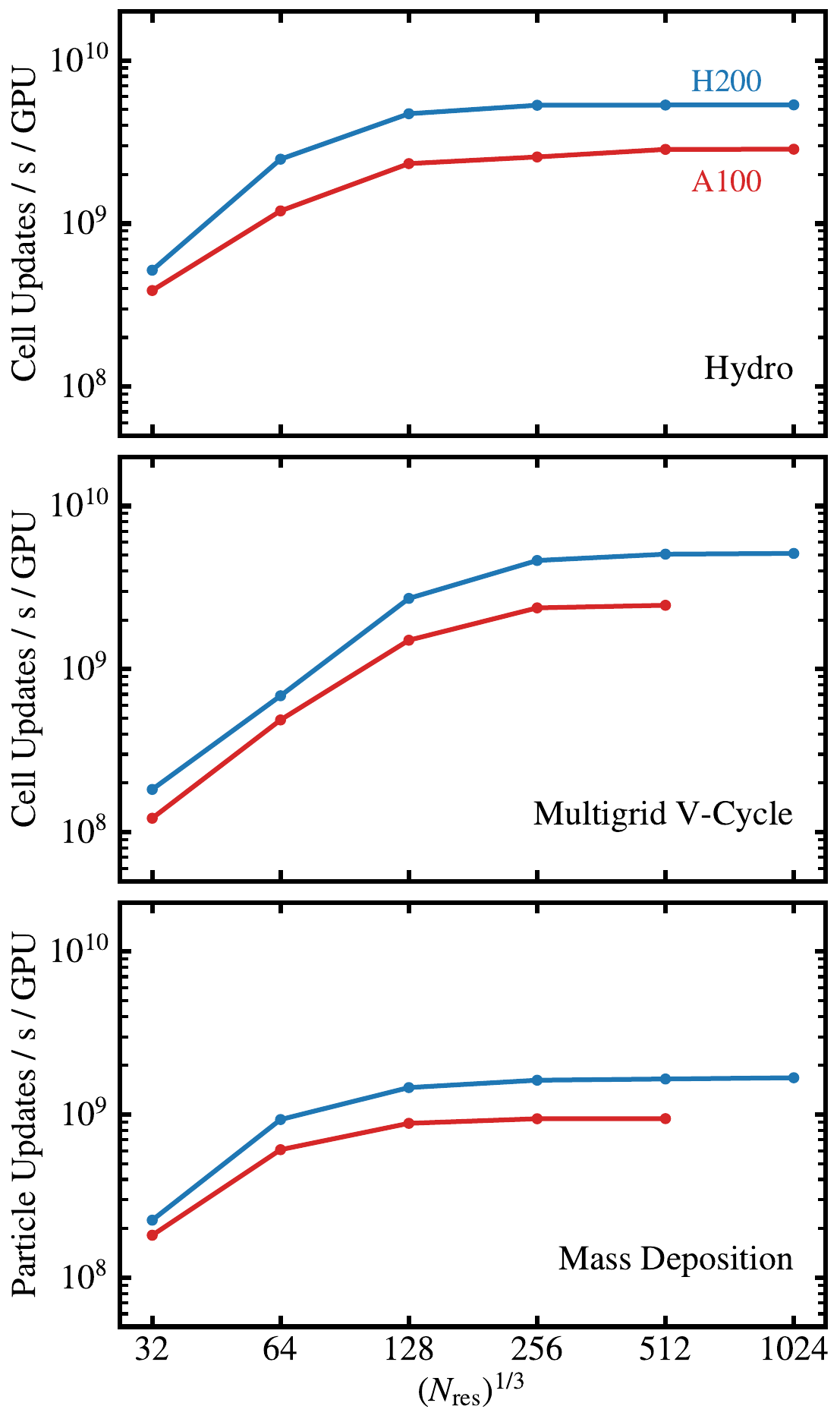}
    \caption{\textbf{Top:} Performance analysis of the hydrodynamics {\it rock} kernel with {\ttfamily nsubgrid=2} and single precision. We show number of cell updates per second per GPU as a function of grid dimensions for the NVIDIA A100 and H200 architectures for the unigrid {\ttfamily sedov} benchmark. \textbf{Middle:} Performance analysis of the MG Poisson solver kernel using the unigrid {\ttfamily coeur} benchmark. We show number of cell updates per second, per GPU and per V-cycle iteration as a function of grid resolution. \textbf{Bottom:} Performance analysis of the particle mass deposition kernel using the unigrid {\ttfamily cosmo} benchmark on A100 and H200 GPUs. The plot shows the number of particle update per second per GPU as a function of the total number of particles.}
    \label{fig:kernel_scalings}
\end{figure}

As shown in the Performance Analysis section, we reach very good performance close to 5 billion cell updates per GPU per second on a H200 for a single V-cycle (see Figure \ref{fig:kernel_scalings}). In practice, we need 4 V-cycles for a residual reduction of $10^{-4}$ (roughly one V-cycle per order of magnitude). The cost of building the MG hierarchy at startup is roughly half a V-cycle.

\section{Portability and Programming Model}

The core acceleration of \textsc{ramses-gpu} is built upon NVIDIA's CUDA Fortran
to access high-performance GPU execution paths directly
from the main simulation codebase. However, to ensure that the code remains
portable across diverse GPU architectures from different hardware vendors (such
as NVIDIA, AMD, and Apple hardware), we avoid as much as possible hard-coding vendor-specific constructs in the main solver loops.

Portability is achieved through a decoupling strategy:
\begin{itemize}
    \item \textbf{Dispatcher and Wrappers:} We implement a dispatcher layer
    alongside carefully designed C--Fortran wrappers (utilizing Fortran's 
    standard \texttt{iso\_c\_binding}). This allows the high-level AMR grid
    management code to interface cleanly with native execution backends.

    \item \textbf{Kernel Translation:} We write or translate target
    computational kernels (e.g., Godunov solvers, CIC deposition, and
    multigrid smoothers) into native CUDA, HIP, and Metal. These kernels are
    directly translated from the existing CUDA Fortran framework, preserving 
    optimized shared memory and register usage patterns.

    \item \textbf{Targeted Execution:} At compile time, the dispatcher selects
    the appropriate back-end implementation based on the destination platform.
    This enables native, high-performance execution on NVIDIA GPUs (via CUDA),
    AMD GPUs (via HIP), and Apple Silicon (via Metal), without the runtime
    overhead or compiler dependence of directive-based paradigms like OpenACC.
\end{itemize}

As an example of this portability strategy, we have implemented a
complete Metal back-end targeting Apple Silicon GPUs. The implementation
introduces three new components, all isolated in a dedicated \texttt{metal/}
directory, leaving the RAMSES Fortran source modified only through
\texttt{\#ifdef \_Metal} preprocessor guards. The first component is a Fortran
module (\texttt{metal\_interface.f90}) containing only ISO C Binding prototypes,
which is entirely backend-agnostic. The second is a thin Objective-C\texttt{++}
host driver (\texttt{metal\_bridge.mm}) that handles device initialization,
persistent buffer allocation via \texttt{MTLBuffer}, command encoding, and
kernel dispatch. The third comprises Metal Shading Language (MSL) compute
kernels directly translated from the CUDA kernels.

The AMR neighbor lookup is handled by the same hash table (with
FNV-1a 64-bit hashing and linear probing). A key adaptation relative to the
CUDA path is that the insertion kernel uses a 32-bit atomic compare-and-swap to
claim hash slots, with the 64-bit Hilbert key written non-atomically after the
slot is secured. This design is required by the absence of 64-bit atomic
operations on Apple GPUs, but is equally valid on HIP and SYCL targets, making
it a portable choice for future back-ends.

Apple Silicon implements a unified memory architecture in which the CPU and GPU
share the same physical DRAM. All device buffers are therefore allocated with
\texttt{MTLResourceStorageModeShared}, so host-to-device and device-to-host
transfers remain within DRAM without any PCIe movement. On discrete AMD or
Intel GPUs targeted by future HIP or SYCL ports, explicit transfer calls will
be required in the host driver, but the Fortran simulation driver and the ISO C
Binding interface layer require no modification. The current implementation
targets single-precision arithmetic throughout (\texttt{NPRE=4}), which is a
hardware constraint of Apple GPUs; back-ends with double-precision support can
use 64-bit types by recompiling with \texttt{NPRE=8} without altering the
kernel logic.

\section{Performance Benchmarks}

\subsection{Sedov Blast Wave Unigrid Test}

We evaluated the core hydrodynamics solver performance using a standard three-dimensional Sedov blast wave benchmark using a pure unigrid configuration. We used for this test the {\it rock} kernel with {\ttfamily nsubgrid=2}. The top panel of Figure~\ref{fig:kernel_scalings} illustrates the rate of cell update per second per GPU on various grid sizes, from $32^3$ to $1024^3$, comparing the NVIDIA A100 and H200 architectures. Note that using single precision, a grid as large as $1024^3$ cells fits into the main GPU memory for both H200 and A100 models. The H200 consistently outpaces the A100, reaching peak performance exceeding 5 billion cell updates per GPU per second as the grid dimensions scale past $N_{\text{cells}} = 128^3$.

\begin{figure}
    \centering
    \includegraphics[width=\columnwidth]{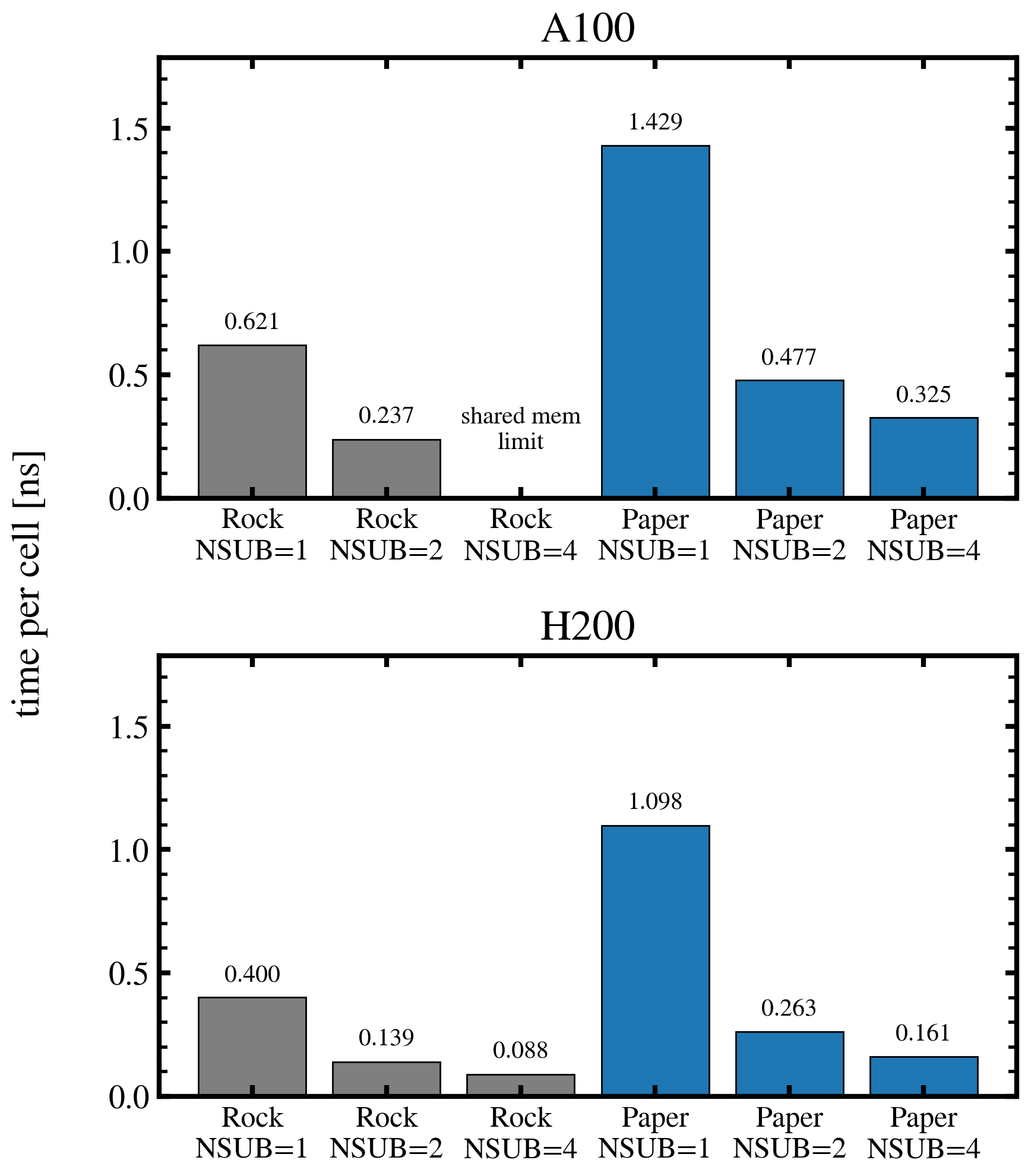}
    \caption{Processing time per cell (in nanosecond) on the A100 and the H200 for the Godunov solver only, showing the efficiency gains of different shared memory kernel designs and different values of \texttt{nsubgrid}. The compilation was performed in single precision and using the {\ttfamily fastmath} option.}
    \label{fig:sedov_cell_time}
\end{figure}

Performance analysis, in terms of processing time per cell, presented in Figure~\ref{fig:sedov_cell_time}, highlights the efficiency gains of different shared memory designs. On the H200, the {\it rock} kernel reaches a whopping speed of 0.09~ns per cell, while the {\it paper} kernel reaches 0.2~ns per cell, both with {\ttfamily nsubgrid=4}. However, on the A100, only the {\it paper} kernel fits in the available shared memory for {\ttfamily nsubgrid=4}. Its performance is equivalent to the {\it rock} kernel with {\ttfamily nsubgrid=2}. Note that in these two cases, the shared memory footprint is around 40~kB for \texttt{nvar = 5} fluid variables. For comparison, we performed the same benchmark on a 96-core Xeon Intel node using MPI reaching 8.9~ns per cell or 112 million cell-updates per second. The H200 delivers a speedup of almost 100x. 

We also report the same benchmark performed on a Mac Book Pro Apple M1 Max with the {\it rock} kernel and \texttt{nsubgrid=1} with performance of 3.3~ns per cell. This is almost 3x faster than the 96-core Intel Xeon node. The H200 is 8x faster than the M1 Max GPU for the same kernel configuration, even though its peak power consumption and price tag are both more than 10x higher. We also performed the same benchmark using 8 MPI ranks on the Apple Book Pro: the M1 Max GPU was 10x faster than the 8 Apple CPU cores. 

\subsection{Sedov Blast Wave AMR Test}

\begin{table*}
\centering
\caption{Detailed execution profiling for the Sedov blast wave test with Adaptive Mesh Refinement, comparing the baseline Intel Xeon 6248R 96-core CPU against a single NVIDIA H200 GPU (FP32 precision).}
\label{tab:sedov_amr}
\begin{tabular}{lrrrrr}
\hline
Step & CPU Time (s) & CPU (\%) & GPU Time (s) & GPU (\%) & Speedup \\
\hline
refine           & 246.886  & 4.9\%   & 66.125   & 15.9\%  & $3.7\times$  \\
cache            & --       & --      & 76.263   & 18.3\%  & --           \\
compute dt       & 96.140   & 1.9\%   & 6.143    & 1.5\%   & $15.7\times$ \\
hydro - set unew & 55.004   & 1.1\%   & 2.510    & 0.6\%   & $21.9\times$ \\
hydro - godunov  & 2599.028 & 52.0\%  & 220.275  & 53.0\%  & $11.8\times$ \\
hydro - set uold & 55.634   & 1.1\%   & 2.520    & 0.6\%   & $22.1\times$ \\
hydro - upload   & 39.000   & 0.8\%   & 23.273   & 5.6\%   & $1.7\times$  \\
flag             & 1901.258 & 38.1\%  & 18.756   & 4.5\%   & $101.4\times$\\
\hline
TOTAL            & 4994.216 & 100.0\% & 415.910  & 100.0\% & $12.0\times$ \\
\hline
\end{tabular}
\end{table*}

\begin{table*}
\centering
\caption{Detailed execution profiling for the cosmological zoom simulation with Adaptive Mesh Refinement, comparing the baseline Intel Xeon 6248R 96-core CPU against a single NVIDIA H200 GPU (FP64 precision).}
\label{tab:cosmo_zoom}
\begin{tabular}{lrrrrr}
\hline
Step & CPU Time (s) & CPU (\%) & GPU Time (s) & GPU (\%) & Speedup \\
\hline
refine              & 432.564   & 1.9\%   & 65.556   & 8.2\%   & $6.6\times$  \\
cache               & --        & --      & 87.796   & 11.0\%  & --           \\
rho                 & 800.344   & 3.4\%   & 185.430  & 23.1\%  & $4.3\times$  \\
poisson             & 19160.896 & 82.5\%  & 401.743  & 50.1\%  & $47.7\times$ \\
grav force          & 385.845   & 1.7\%   & 8.985    & 1.1\%   & $42.9\times$ \\
particle - kickdrift& 1585.721  & 6.8\%   & 20.121   & 2.5\%   & $78.8\times$ \\
compute dt          & --        & --      & 5.971    & 0.7\%   & --           \\
flag                & 851.116   & 3.7\%   & 25.086   & 3.1\%   & $33.9\times$ \\
\hline
TOTAL               & 23234.169 & 100.0\% & 801.637  & 100.0\% & $29.0\times$ \\
\hline
\end{tabular}
\end{table*}

To validate performance under AMR conditions, we also run the Sedov blast wave test using adaptive mesh refinement. Table~\ref{tab:sedov_amr} presents the execution timing comparison between the H200 Nvidia GPU and a typical Intel Xeon multi-core platform. We used the rock kernel and \texttt{nsubgrid=1} with \texttt{levelmin=8} and \texttt{levelmax=10} for the \texttt{sedov} benchmark until time $t=5\times 10^{-2}$. 

The number of octs ($2\times 2\times 2$ cell blocks) was 2,097,152 at $\ell=8$, 1,602,359 at $\ell=9$, and 4,116,151 at $\ell=10$ at the end of the simulation. 
We first see that for both architectures, the cost of doing AMR is roughly slowing down the execution by a factor of 2. The overall speed up between a single H200 and 96-core Xeon is 12x. We can see the non-negligible cost (12~\%) of handling ghost zones on the GPU.

\subsection{Core Collapse Unigrid Test}

The MG Poisson solver can be validated using a simple classical pure gas self-gravity problem \citep{Boss1979}. It features the gravitational collapse of a dense molecular core into a binary stellar system. Performance analysis is reported in the middle panel of Figure~\ref{fig:kernel_scalings} and reveals a maximum throughput per MG V-cycle of around 5 billion cell updates per GPU per second. Saturation is reached for grid sizes larger than $128^3$. The cost of building-up the MG hierarchy and initializing the solution amounts to roughly half a V-cycle. For typical astrophysical use cases, users require the residual to converge to a relative residual norm of 4 or 5 orders of magnitude, which corresponds to 4 or 5 V-cycles. Overall, for a unigrid Cartesian mesh, a complete MG solve of the Poisson equation proceeds at 1 billion cell update per GPU per second.

We also report the same benchmark performed on a Mac Book Pro Apple M1 Max with a performance of 4.5~ns per cell per V-cycle. The H200 is 20x faster than the M1 Max GPU for this kernel. We also performed the same benchmark using 8 MPI ranks on the Apple Book Pro: the M1 Max GPU was 16x faster than the 8 Apple CPU cores.

\subsection{Cosmological Simulation Unigrid Test}

This test is used to benchmark the N-body solver, combining the particle pusher, the mass deposition scheme, the force interpolation scheme, and the gravity solver. The most time consuming step is the gravity solver, followed by the Cloud-in-Cell mass deposition scheme. Figure~\ref{fig:kernel_scalings} presents the performance analysis of the CIC mass deposition scheme for a simulation with $N_{\rm part}=N_{\rm grid}$ and $N_{\rm part}$ ranging from $32^3$ to $1024^3$. Our best performance is reached for $N_{\rm part}>64^3$ at around 5 billion particle depositions per GPU per second on the H200. Note that we used for this test the Atomic-Add mass deposition scheme, as it performs significantly better than the Prefix-Sum alternative in this case.

We have also ported the Atomic-Add CIC kernel to the Apple Metal GPU. The overall mass deposition scheme amounts to 21~ns per particle, which is significantly slower than out best H200 performance. The mass deposition itself is only 7~ns, while the particle Hilbert key sorting costs 14~ns. This approach suffers from not being able to use the highly optimized NVIDIA CUB radix sort. Our Metal GPU execution is nevertheless 6x faster than the 8 core MPI run on the Apple M1 Max.

\subsection{Cosmological Zoom-In Simulation AMR Test}

Finally, to test the Particle-Mesh code and the gravity solver under high density contrast conditions and using a deep AMR grid with adaptive time stepping, we execute a high-resolution cosmological zoom-in pure dark matter simulation. The CPU baseline run (executed on an Intel Xeon 6248R 96-core processor) is directly compared against a single NVIDIA H200 GPU running both in double precision (FP64).

The total elapsed wall-clock time drops from 23,082.9 seconds on the 96-core CPU platform to just 802.6 seconds on the H200 GPU, achieving an overall execution speedup of approximately $28.7\times$. Detailed profiling breaks down the execution time per computational step in both architectures in Table~\ref{tab:cosmo_zoom}. The CPU code is dominated by the Poisson solver, which suffers from the latency of multiple MPI communications. This could be optimized up to 2x at the expense of using more buffer memory. Overall, taking advantage of the data being resident in the GPU memory, porting the code to GPU allows us to consistently get a speedup of at least 10x in realistic setups.

\section{Ported Modules and Future Directions}

Beyond the core solvers described in this paper (hydro, gravity, particles and AMR), several key additional physics modules have already been successfully ported to the GPU using the same CUDA Fortran and Apple Metal frameworks:
\begin{itemize}
    \item \textbf{Gas Physics:} Equilibrium chemistry cooling and various polytropic Equations of States.
    \item \textbf{Magnetic Fields:} Ideal and non-ideal Magneto-hydrodynamics (MHD) solvers using constrained transport.
    \item \textbf{Star Formation:} Various subgrid star formation recipes and support for star particles.
    \item \textbf{Stellar Feedback:} Thermal and mechanical supernova (SN) feedback.
    \item \textbf{Self-Gravity:} A Fast Multipole Method (FMM) gravity solver for non-periodic boundaries.
\end{itemize}

Active development is still ongoing on porting the remaining segments of the \textsc{ramses} ecosystem, including MPI for multi-GPU capabilities, sink particles, supermassive black hole (SMBH), the PHEW clump finder, Active Galactic Nuclei (AGN) feedback models, and dedicated routines tracking dust grains and passive tracer particles.

\section{Conclusions}
We have developed and tested \textsc{ramses-gpu}, a GPU CUDA and Metal port of the cell-by-cell Adaptive Mesh Refinement astrophysical simulation code \textsc{ramses}. Using CUDA Fortran and optimized kernel implementations alongside specialized device libraries such as CUB, we handle the complex, dynamic grid structures of AMR with minimal overhead. The resulting code delivers major performance speedups for both idealized hydrodynamics tests and realistic cosmological production runs, paving the way for next-generation galaxy formation simulations on exascale heterogeneous computing platforms. 

It is worth mentioning that results produced by the GPU versions of \textsc{ramses} are identical to the CPU version if both versions are compiled with the same floating point accuracy. Results might diverge slowly in the case of non-linear dynamics and nondeterministic parallel updates between the GPU and the CPU MPI versions. The random number generator used in the GPU version (cuRAND) in the star formation routine also differs significantly from the CPU version, causing the results to diverge more, but they are still consistent on a statistical level. 

The next major step for us will be to develop the multi-GPU implementation, probably based on the MPI library.

\section*{Code Availability}
The source code for the simulations presented in this work is freely available in the \textsc{mini-ramses} repository on Bitbucket at https://bitbucket.org/rteyssie/mini-ramses.

\section*{Acknowledgements}
This work was supported by Princeton University and NVIDIA during the Princeton Open Hackathons 2025 and 2026. We acknowledge support from Princeton Research Computing via partial funding of a Research Software Engineer. This material is based upon work supported by the National Science Foundation (NSF) and the
U.S.-Israel Binational Science Foundation (BSF) under Award Number 2406558 and Award Title ``The Origin of the Excess of Bright Galaxies at Cosmic Dawn''. This work was supported in part by the U.S. Department of Energy SLAC Contract No.DE-AC02-76SF00515. JS acknowledges and is grateful for financial support from the Fannie \& John Hertz Foundation. The simulations presented in this article were performed on computational resources managed and supported by Princeton University’s Research Computing.

\bibliographystyle{rasti}
\bibliography{references}

\label{lastpage}
\end{document}